\documentclass[11pt]{article} 

\usepackage[utf8]{inputenc} 

\usepackage{graphicx} 
\usepackage{amsmath,amsthm,amsfonts}

\usepackage{mathpazo}
\usepackage{eulervm}
\usepackage[vmargin=3.5cm,hmargin=4.5cm]{geometry}

\newcommand{\real}{\mathbb{R}}

\newcommand{\half}{\frac{1}{2}}

\newtheorem*{theorem}{Theorem}

\newtheorem{claim}{Claim}

\title{Conductance and Eigenvalue}

\author{Girish  Varma\\ girish@tcs.tifr.res.in \\  \\ School of Technology and Computer Science\\Tata Institute of Fundamental Research, Mumbai}

\begin{document}

\maketitle

\begin{abstract}
We show the following.
\begin{theorem}
Let $M$ be an finite-state ergodic time-reversible Markov chain with
transition matrix $P$ and conductance $\phi$. Let $\lambda \in (0,1)$ be
an eigenvalue of $P$. Then,
$$\phi^2 + \lambda^2 \leq 1$$
\end{theorem}
This strengthens the well-known~\cite{HLW,Dod84, AM85, Alo86, JS89}
inequality $\lambda \leq 1- \phi^2/2$.  We obtain our result by a
slight variation in the proof method in \cite{JS89, HLW}; the same
method was used earlier in \cite{RS06} to obtain the same inequality
for random walks on regular undirected graphs.
\end{abstract}

A \emph{Markov chain} is a sequence of random variables $\{X_i\}_{i
\geq 1}$ taking values in a finite set such that
\begin{eqnarray*}
\Pr[X_t = i \mid X_{t-1} = j, X_{t-2} =x_{t-2}, \cdots, X_0 = x_0] 
 = \Pr[X_t = i \mid X_{t-1} = j].
\end{eqnarray*}
Let the state space of the Markov chain be $[n]$ and let $P= (P_{ij})$
be its $n \times n$ transition matrix: $P_{ij} = \Pr[X_t = i \mid X_{t-1}
= j]$. We will assume that the Markov chain is ergodic, that is, irreducible( for every pair of states $i,j \in [n]$, $P^s_{ij} > 0$ for
some $s$) and aperiodic(for any state $i\in [n]$, $ \text{gcd} \{s\ :\ P^s_{ii} > 0\} = 1$). Then, the Markov chain has a unique stationary distribution $\pi$: $P\pi = \pi$.
We say that the Markov chain is time-reversible if it satisfies
the following {\em detailed balance condition:}
\begin{equation}
\label{det-bal}
\forall i,j \in [n],~  P_{ij} \pi_j =  P_{ji}\pi_i
\end{equation}
All Markov chains considered in this note will be assumed to be finite-state ergodic and time-reversible. The \emph{conductance} of a Markov chain with state space $[n]$ is defined to be
$$\phi = \min_{S \subset [n]: \sum_{i \in S} \pi_i \leq 1/2}
\frac{\sum_{i\in S, j \notin S}  P_{ji} \pi_i}{\sum_{i \in S} \pi_i}$$

The following theorem plays a central role in the theory of rapidly mixing
Markov chains.
\begin{theorem}[\cite{JS89}]
Let $\lambda < 1$ be an eigenvalue of the transition matrix of an ergodic time-reversible Markov chain with conductance $\phi$. Then,
 $\lambda \leq 1 - \frac{\phi^2}{2}$.
\end{theorem}
In this note we strengthen this inequality slightly.
\begin{theorem}
Let $\lambda \in (0,1)$ be an eigenvalue of the transition matrix of
an ergodic time-reversible Markov chain with conductance $\phi$. Then,
$$\phi^2 + \lambda^2 \leq 1$$
\end{theorem}
Such an inequality was derived by Radhakrishnan and Sudan~\cite{RS06}
for the special case of random walks on regular undirected
graphs. The purpose of this note is to show that their arguments
(which were a slight variation on the arguments in~\cite{JS89, HLW})
apply to finite-state ergodic time-reversible Markov chains as well.

\begin{proof}
 Let $\pi$ be the stationary distribution of the chain with transition matrix $P$. Let $f,g\in \real^n$. We will be thinking of $f,g,\pi$ as vectors in $\real^n$. Let 
 $$\langle f, g \rangle = \sum_{i\in [n]} f_i \pi_i g_i $$
 and $||f|| = \sqrt{\langle f, f \rangle}$.
 $f$ is said to be \emph{proper} if 
  $$f \neq 0 ~\text{ and }~ \forall i \in [n],~ f_i \geq 0 ~\text{ and } \sum_{i\in[n]: f_i >0} \pi_i \leq \half $$
 
 We have the following two claims.
 
 \begin{claim}
\label{claim-conductance}
For any proper $f$, 
\begin{equation}
\label{eq:conductance}
\phi^2 || f ||^4 \leq || f ||^4 - \langle f ,P^T f \rangle^2 
\end{equation}
\end{claim}
 
\begin{claim}
\label{claim-laplacian}
For $\lambda \in (0,1)$, there exists a proper $f$ such that
 \begin{equation}
\label{eq:laplacian}
\langle f, P^T f \rangle   \geq \lambda || f ||^2
\end{equation}
\end{claim}
Using (\ref{eq:conductance}) and (\ref{eq:laplacian}), we obtain
$$\phi^2 || f ||^4 \leq || f ||^4 - \lambda^2 || f ||^4 $$
from which the theorem follows. \\
\textit{Proof of Claim \ref{claim-conductance}}.
Permute the co-ordinates of $f$ such that $f_1 \geq f_2 \geq \cdots \geq f_r > 0 \text{ and } f_{r+1} = \cdots = f_n = 0$. (Note that $\sum_{i\in [r]} \pi_i \leq 1/2$.) We show that 
$$\phi^2 || f ||^4  \leq \left[\sum_{i<j} P_{ij}\pi_j(f_i^2 -f_j^2)\right]^2 \leq || f ||^4 - \langle f ,P^T f \rangle^2 $$

To see the first inequality, we observe that
\begin{align*}
\sum_{i<j}  P_{ij} \pi_j (f_i^2 -f_j^2) & =  \sum_{i<j}  P_{ij} \pi_j \sum_{i\leq k <j} (f_k^2 -f_{k+1}^2) \\
	& =  \sum_{k \in [r]} (f_k^2 - f_{k+1}^2) \sum_{i\in[k], j \notin [k]} P_{ij} \pi_j \\
	& \geq  \phi \sum_{k \in [r]} (f^2_k - f^2_{k+1})\left(\sum_{i \in [k]} \pi_i \right) \\
	& = \phi \sum_{k \in [r],i \in [k]} (f^2_k - f^2_{k+1}) \pi_i \\
	& =  \phi \sum_{i\in [r]} \pi_i f^2_i \\
	& = \phi || f  ||^2
\end{align*}
Secondly	
\begin{align*}
\sum_{i<j}  P_{ij} \pi_j (f_i^2 -f_j^2)  & = \sum_{i<j} \sqrt{ P_{ij} \pi_j } (f_i -f_j) \sqrt{ P_{ij} \pi_j } (f_i +f_j) \\
	& \leq   \left[  \sum_{i<j}  P_{ij} \pi_j (f_i -f_j)^2  \sum_{i<j}  P_{ij} \pi_j (f_i +f_j)^2\right]^{\half}  \\  & \hspace{1.5 cm} \text{(by the Cauchy-Schwarz inequality)} 
	\end{align*}	
	The calculations up to this point are identical to those in \cite{JS89,HLW}; the calculations below are similar to those in \cite{RS06}.
	\begin{eqnarray*}
		& =&   \left[  \sum_{i<j}  P_{ji} \pi_i  (f_i^2 +f_j^2 -2f_i f_j) \sum_{i<j} P_{ji} \pi_i  (f_i^2 +f_j^2 +2f_i f_j)\right]^{\half} \\
	 &=&  \left[  
	\left( \sum_{ij}  P_{ji} \pi_i  f_i^2 - \sum_{ij}  P_{ji} \pi_i  f_i  f_j \right) 
	  \left( \sum_{ij}  P_{ji} \pi_i  f_i^2 + \sum_{ij} P_{ji} \pi_i  f_i  f_j -2\sum_i  P_{ii} \pi_i f_i^2 \right) 
	\right]^{\half} (\text{using } \ref{det-bal})\\
	& \leq & \left[ \left( \sum_{i} \pi_i  f_i^2 - \sum_{ij} P_{ji} \pi_i  f_i  f_j \right) 	  \left( \sum_{i} \pi_i  f_i^2 + \sum_{ij} P_{ji} \pi_i  f_i  f_j \right) \right]^{\half} \\
	  & = & \left[ \left( \sum_{i}  \pi_i f_i^2 \right)^2 - \left(\sum_{ij} P_{ji} \pi_i f_i  f_j \right)^2 \right]^{\half} \\
	& = & \left[ ||f||^4 -  \langle f, P^T f \rangle^2 \right]^{\half}
	\end{eqnarray*}
\textit{Proof of Claim \ref{claim-laplacian}}.
Let $g\in \real^n$ be a right eigenvector of $P^T$ with eigenvalue $\lambda \in (0,1)$ . We may assume  $ \sum_{i: g(i) > 0} \pi_i \leq \half$ (otherwise consider $-g$). By renaming the co-ordinates we may assume that $g_1 \geq g_2 \geq \cdots \geq g_r > 0 \geq g_{r+1} \geq g_{r+2} \geq \cdots \geq g_n$. Let $f$ be such that $f_i = g_i$ for $i\in[r]$ and $0$ otherwise. Then
 $$\forall i \in [r],~ (P^T f)_i \geq (P^T g)_i = \lambda g_i = \lambda f_i$$
 Then, $\langle f, P^T f \rangle = \sum_{i\in[r]}\pi_i f_i(P^T f)_i  \geq \lambda \sum_{i \in [r]} \pi_i f_i^2 = \lambda ||f||^2$.

\end{proof}
\bibliographystyle{plain}
\bibliography{grv}

\end{document}